# Coupling nitrogen vacancy centers in silicon carbide to nanophotonic resonators

Ivan Zhigulin[†,1,2], Konosuke Shimazaki[†,1,2], Samuel M. Stephens[1,2], Angus Gale[1,2], Karin Yamamura[1,2], Hark Hoe Tan[3], Igor Aharonovich[*,1,2] and Mehran Kianinia[*,1,2]

[1] School of Mathematical and Physical Sciences, University of Technology Sydney, Ultimo, New South Wales 2007, Australia
[2] ARC Centre of Excellence for Transformative Meta-Optical Systems, University of Technology Sydney, Ultimo, New South Wales 2007, Australia
[3] ARC Centre of Excellence for Transformative Meta-Optical Systems, Department of Electronic Materials Engineering, Research School of Physics, The Australian National University, Canberra, ACT 2600, Australia

† These authors contributed equally to this work.
* To whom correspondence should be addressed: I.A. Igor.Aharonovich@uts.edu.au, M.K. Mehran.Kianinia@uts.edu.au

## Abstract

*Silicon carbide (SiC) is a promising platform for scalable quantum technologies owing to its well-established, wafer-scale industrial processing. SiC also hosts a variety of optically active color centres including the nitrogen vacancy defect that has a spin-triplet ground state. However, strong phonon coupling in the infrared range limits photon extraction from these defects. Here, we use nanophotonic structures, specifically micro-pillar and micro-disk resonators, to enhance optical collection and spin-readout. The micro-pillar geometry yields a 4-fold increase in photon collection, accompanied by a 2.4-fold reduction in spectral noise in optically detected magnetic resonance measurements. Consequently, the magnetic field sensitivity is improved by 24%. The large mode volume of the micro-disk supports resonances spanning 1150-1250 nm, enabling broadband coupling to nitrogen vacancy emission lines. Our results demonstrate that fabrication of scalable photonic structures efficiently improves performance of silicon carbide color centers for integrated quantum light generation and sensing.*

## Introduction

Novel quantum architectures enable applications ranging from nanoscale sensing and single-photon communication to photonic qubit processing[1,2]. These can be realised using quantum systems as building blocks, in particular, solid-state color centers with optically interfaced quantum states[3–5]. The choice of the host material is then directly linked to the practicality and scalability of these platforms, providing wafer-scale fabrication and compatibility with complementary metal-oxide-semiconductor (CMOS) processing[6–10].

Silicon carbide (SiC) offers mature industrial fabrication infrastructure while hosting a wide range of spin-active color centers, including several vacancy-related defects with ground-state spin-triplet configurations that enable optically detected magnetic resonance (ODMR) and coherent spin manipulation[11]. Among these, nitrogen vacancy (NV) centers have emerged as promising candidates due to their deterministic generation[12], near-infrared spectral range[13], room-temperature spin coherence at the single emitter level[14], and demonstrated applications in radio-frequency sensing[15]. Despite these advantages, NV centers are limited by strong phonon coupling, which broadens the emission spectrum and significantly reduces photon extraction efficiency. These limitations constrain their performance as single-photon sources and spin sensors. Integration into scalable on-chip nanophotonic structures provides a pathway to overcome these challenges by enhancing photon collection and improving signal-to-noise ratios[8,16].

Among various nanophotonic approaches, μ-scale geometries offer a more practical path as opposed to resonant nanocavities that require stringent fabrication and spectral alignment. In particular, μ-Pillars efficiently redirect emitted light into a cone, increasing the effective numerical aperture (NA) of the collection system. μ-Disks, on the other hand, support multiple optical modes using a semi-suspended geometry, avoiding complete undercutting that is often challenging for fully suspended structures (such as photonic crystal cavities). Their design enables broadband enhancement of photon extraction, making them highly suitable for coupling with multiple NV configurations in 4H-SiC.

Here, we generate NV centers in 4H-SiC μ-structures, specifically μPillars and μ-Disks, designed to enhance optical extraction and spin readout efficiency. Compared to NV centers in bulk SiC, our devices enable observation of non-classical emission without the need for high numerical aperture (1.2 ≥ NA) oil-immersion objectives, which have previously limited such measurements to room-temperature operation[12,14,17]. Thus, we show μ-Pillars to be a suitable platform for conducting advanced quantum experiments. Additionally, higher signal-to-noise ratios from both μ-Pillars and μ-Disks showed significant improvement of NV sensing performance. Our results advance telecom quantum emitters in scalable on-chip devices for quantum light generation and spin control in quantum photonics.

**Results and Discussion**

The experimental concept of this work is illustrated in Figure 1a as a three-dimensional render schematic. High-purity 4H-SiC was commercially sourced and used to fabricate monolithic μ-Pillars and μ-Disks. Following fabrication, generated NV centers were optically addressed using a continuous-wave laser, while a suspended copper wire positioned above the structures delivered microwave (MW) fields for coherent spin control.

Crystal structure of 4H-SiC consists of offset hexagonal lattices of silicon and carbon with ABCB stacking, giving rise to two nonequivalent lattice sites, denoted h (quasi-hexagonal) and k (quasi-cubic). As a result, four distinct NV configurations are possible within this lattice. These are illustrated in the inset of Figure 1a, where hh and kk are referred to as axial NVs, while hk and kh are referred to as basal NVs. All configurations exhibit the same electronic structure, shown in Figure 1b. However, the zero-phonon line (ZPL) energy depends on the specific NV configuration, accounting for variations in atomic separations between the nitrogen and vacancy sites.

The NV centre ground state forms a spin triplet consisting of $|m_s = 0\rangle$ and $|m_s = \pm 1\rangle$ sublevels.

Application of resonant MW fields drives transitions between these sublevels, followed by spin-preserving optical excitation into the excited state. The electron then can either radiatively decay back to the ground state or undergo non-radiative intersystem crossing (ISC) into a metastable singlet state, allowing for optical readout of spin transitions. Owing to the distinct atomic environments of each NV configuration in 4H-SiC, the zero-field splitting (ZFS) parameters defining MW resonance frequencies exhibit variations for each NV[12].

To generate NV centers, SiC is first implanted with nitrogen ions using an ion implanter. Afterwards, it is necessary to anneal the sample at 1000 °C in high vacuum to promote vacancy diffusion and formation of nearest-neighbour NV centers. However, during this step, a competing defect formation also occurs, primarily the aggregation of two vacancy sites that form a divacancy (VV) center. Emission from VV exhibits phonon sideband that partially overlaps with NV ZPLs and involves spin transitions at nearby resonances to NV, contributing to ODMR spectra. Influence of VV defects on the measured optical and spin results are discussed later in this manuscript.

NV generation in μ-Pillars was performed prior to nanofabrication using a 200 keV broad nitrogen ion beam with an implantation dose of $1 \times 10^{12}$ ions/cm$^2$. Simulations of ion penetration and stopping range showed the nitrogen ion profile peaks at depth of ~310 nm (Figure 1c). For μ-Disks, implantation was carried out post-fabrication at a reduced energy of 30 keV. Because of the suspended geometry of disk structures, this approach enabled controlled, shallow implantation depths matched to the disk thickness. Using doses of $2.14 \times 10^{12}$ and $5.35 \times 10^{12}$ ions/cm$^2$ resulted in a profile peak of ~60 nm.

Following defect generation, cryogenic-temperature photoluminescence (PL) spectroscopy was employed to probe the formation of all NV configurations in each sample, as room-temperature NV emission is strongly broadened by phonon coupling (see Figure SI1). Measurements were also performed using a 1050 nm optical pump to minimise excitation of divacancy (VV) centers. High-resolution representative cryogenic PL spectrum using 1200 l/mm grating is shown in Figure 1d. All four NV ZPLs are observed with peaks centred at 1175.7 nm, 1179.1 nm, 1222.3 nm, and 1242.2 nm corresponding for NV$_{kh}$, NV$_{hh}$, NV$_{kk}$, and NV$_{hk}$, respectively. Variations in the relative intensities are attributed to factors such as local defect density and emitter dipole orientation. Notably, the NV generation method also forms tungsten-silicon (W$_{Si}$) centers[15,18], whose emission line is observed in the spectrum at 1172.2 nm. Having confirmed the formation and optical accessibility of all NV configurations in 4H-SiC, we proceed to their integration into nanophotonic structures.

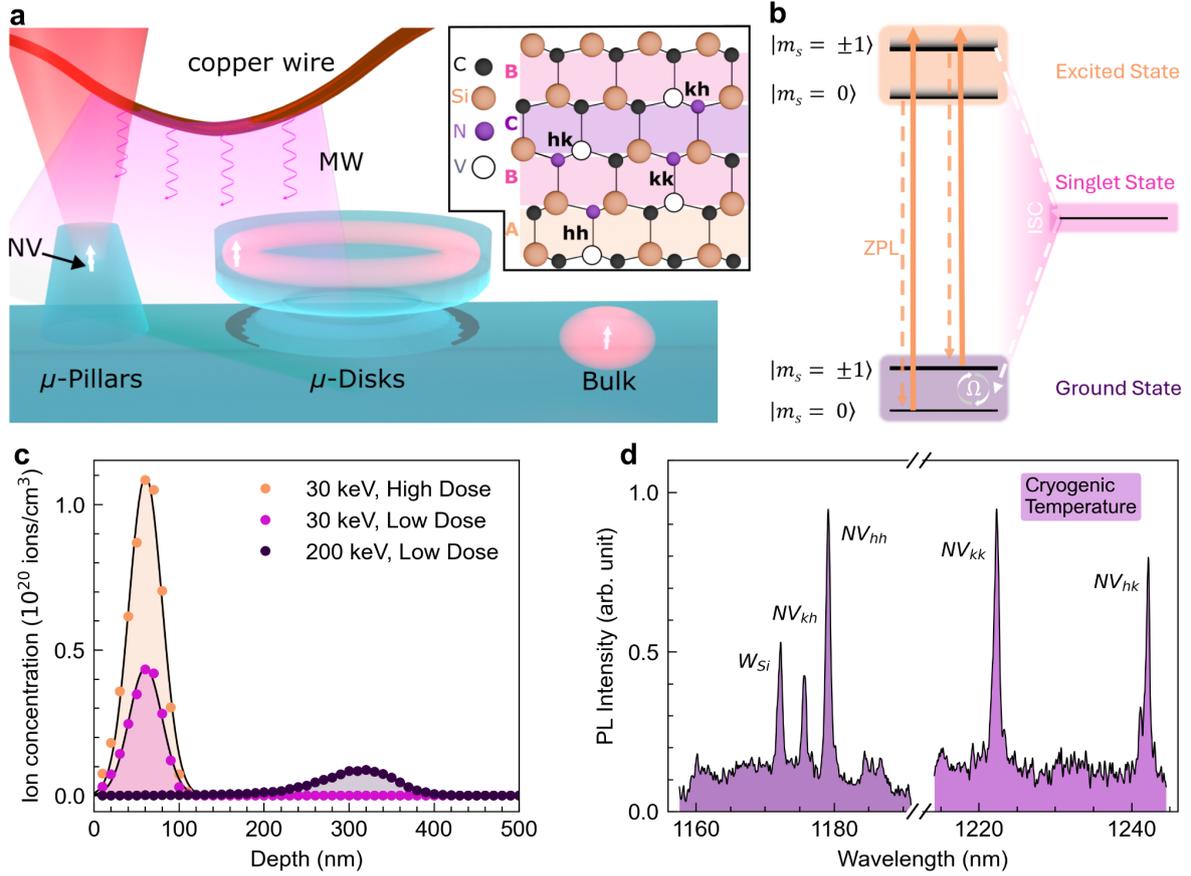

*Figure 1: Experimental setup and NV generation. (a) Schematic illustration of the experimental setup showing NV emission fields in SiC photonic structures and bulk. Inset: Atomic structure configurations of NV centers in 4H-SiC. (b) Energy diagram of NV centers in 4H-SiC, where ZPL denotes zero-phonon line photon emission and ISC is intersystem crossing of an electron into the singlet state. (c) Simulation of penetration depth distributions of $N^+$ ions using 30 keV ($2.14(10^{12})$ and $5.35(10^{12})$ ions/cm$^2$) and 200 keV ($1(10^{12})$ ions/cm$^2$) acceleration voltages. (d) Emission spectra of NV centers at cryogenic temperature showing $NV_{kh}$, $NV_{hh}$, $NV_{kk}$, and $NV_{hk}$ lines at 1175.7 nm, 1179.1 nm, 1222.3 nm, and 1242.2 nm, respectively.*

Fabricated structures are shown in Figure 2a,d as scanning electron microscope (SEM) images. To visualize the optical effect of the nanostructures, we simulated the electric field distribution using the finite-difference time-domain (FDTD) method via Ansys Lumerical. Simulation geometries were constructed directly from the corresponding SEM images to visualise electric fields in the fabricated structures. NV centers were modelled as electric dipole emitters within the wavelength range of 1.1 to 1.5 µm. Dipole orientation was set to be either parallel to the c-axis of 4H-SiC (axial, $NV_{hh}$ and $NV_{kk}$) or at 71° relative to the c-axis (basal, $NV_{kh}$ and $NV_{hk}$).

Comparison of bulk and µ-Pillar simulations confirms effective enhancement of emission outcoupling for the latter. Figure 2b,c show the electric field distribution at the wavelength of 1.178 µm on the x-z plane containing the dipole. Both basal and axial NV dipole orientations have significantly enhanced field outcoupling into free space collection cones, suggesting improvement of the effective N.A. of the collection system. When both basal and axial NV centers exist, the outcoupling power of the µ-Pillar structure is ~10 times greater compared to the bulk for an objective lens with N.A. of 0.81.

For the µ-Disk geometry, whispering-gallery mode (WGM) is observed in both the electric field distribution and the simulated spectrum. Figure 2e shows the electric field distribution on the x-y plane containing the dipole. The field is confined along the disk perimeter and forms a standing-wave pattern characteristic of WGM in both axial and basal coordination. WGMs are generally classified by three mode indices ($m$, $p$, $q$), corresponding to the azimuthal, radial, and vertical mode numbers, respectively. In our calculation model, rotational symmetry about the $z$ axis is preserved, whereas inversion symmetry about the $x$ axis is not. Consequently, only the azimuthal mode number can be unambiguously identified. We determine $m \approx 34$ by counting the intensity maxima along the disk circumference in $|E|^2$. It is also noted that, in Figure 2e, we observed multiple intensity lobes in the radial direction, indicating the excitation of higher-order spatial modes. Such behavior is also consistently observed in the lateral electric field distribution (Figure SI2), confirming that several WGM-like modes are simultaneously excited in the μ-Disks.

Figure 2f shows the simulated emission spectra for axial and basal NV orientations. Multiple sharp resonances attributable to WGMs are resolved. Small baseline oscillations are also visible and arise from the finite time window used in the Fourier transform of the time-domain window. Taken together, these results confirm that our μ-Disk geometry supports WGMs in the infrared region corresponding to the emission of SiC NV centers.

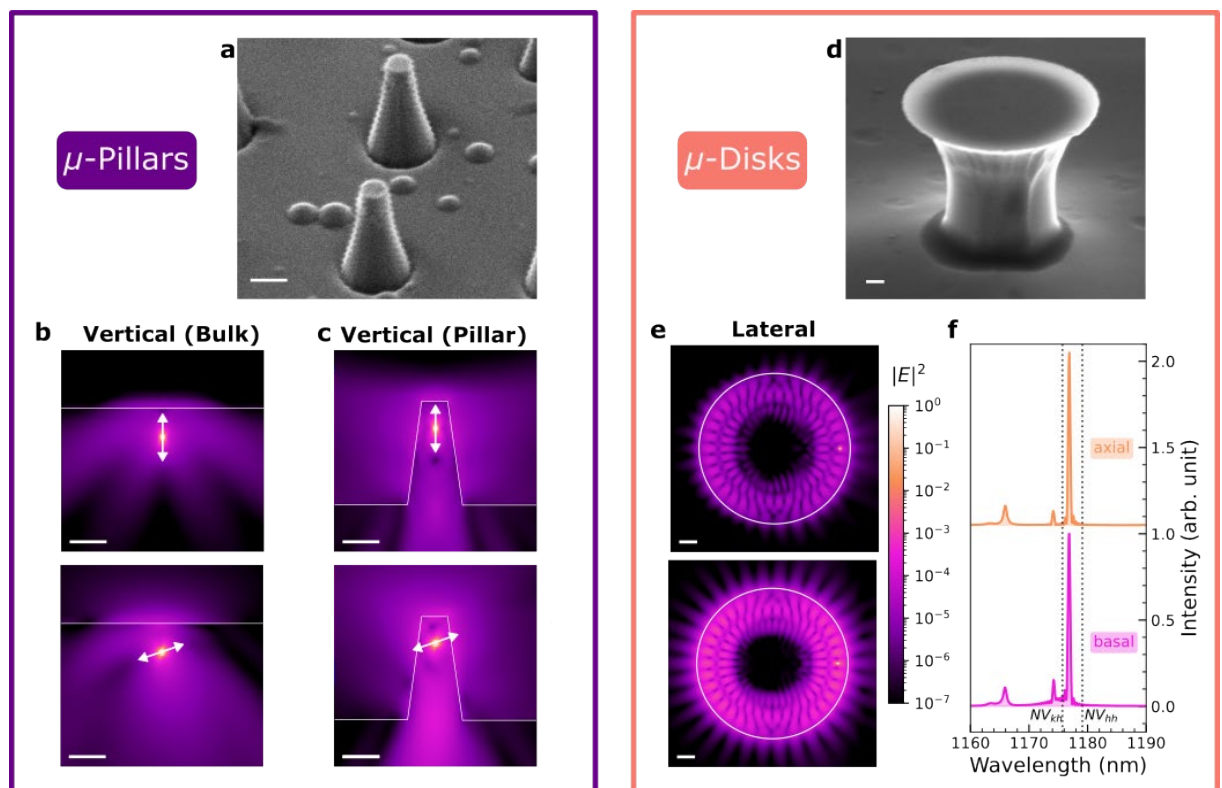

*Figure 2: SEM images and FDTD simulation results of the fabricated 4H-SiC μ-Pillars and μ-Disks hosting NV centers. (a), (d) SEM images of representative μ-Pillar and μ-Disk structures (scale bar: 500 nm). (b), (c), (e) Simulated electric field distributions for axial-oriented (top) and basal-oriented (bottom) NV dipoles. (b) Vertical field distribution in bulk 4H-SiC. (c) Vertical field distribution for an NV center in a 4H-SiC μ-Pillar. (e) Lateral field distribution for an NV center in a 4H-SiC μ-Disk. (f) Simulated optical mode spectrum of a 4H-SiC μ-Disk excited by basal (top, orange) and axial (bottom, pink) dipole sources. Dashed lines correspond to NV ZPL emissions. All field distribution panels have a scale bar of 500 nm.*

In the following sections, we characterize the generated NV centers in SiC photonic μ-structures. To begin, the samples were first cooled to cryogenic temperature (~12 K) and probed using scanning confocal microscopy operating in reflection (details of measurements are provided in the Methods section). Figure 3a shows PL spectra from NV centers in two separate μ-Pillars. Strong ZPL peaks are observed at 1177.5 nm, 1180.5 nm, and 1241.8 nm, corresponding to the $NV_{kh}$, $NV_{hh}$, and $NV_{hk}$ configurations, respectively. Additional spectra from other μ-Pillars are shown in Section 3 of the Supplementary Information. Each μ-Pillar hosted between one and three NV configurations; while the $NV_{hk}$ configuration was not observed in this sample, although the statistical variation is limited. Importantly, in addition to NV emission, strong presence of VV is persistent in all measured μ-Pillars and bulk SiC. This resulted in unavoidable phonon sideband emissions that spectrally overlap with NV ZPLs (Figure SI3), effectively elevating the background level beneath NV PL.

Despite this background contribution, the μ-Pillar geometry provides sufficient enhancement to observe first photon antibunching from NV centers at cryogenic temperature. At low optical excitation powers, we used a fibre-based Hanbury Brown-Twiss interferometer to acquire a bi-directional autocorrelation histogram, shown in Figure 3b. The dataset is measured from a $NV_{hh}$ center in Pillar 1. Prior to the correlation measurement the PL signal was spectrally filtered using a combination of optical filters to isolate the narrowband $NV_{hh}$ emission (Figure SI4). The histogram reveals a pronounced antibunching dip at short delay times ($\tau \leq 1$ ns) and is fit with a two-level second-order autocorrelation model (described in Supp. Section 3) that yields $g^{(2)}(0) = 0.7$. As discussed earlier, phonon-sideband emissions of VV and $W_{si}$ defects overlap with $NV_{hh}$ ZPL even within the filtered spectral window, thus elevating the recorded antibunching value. Minimising contributions of these background emitters would significantly lower $g^{(2)}(0)$ value, potentially falling below the single-photon threshold of 0.5. This highlights opportunities for optimisation of NV generation aimed at minimising formation of VV and other centers primarily during annealing step, either via lowered temperatures[19] or altered cooldown process[20]. Furthermore, multi-wavelength excitation schemes could be used to quench PL of VV through conversion into the non-luminescent positive charge state[21–24].

We confirmed the presence of WGMs in the μ-Disks by measuring the emission spectrum from a disk without ion implantation, as shown in Figure 3c (lowest dataset). Several narrow peaks are visible, indicating the formation of whispering-gallery modes in the fabricated structures. To verify the successful formation of NV centers, PL spectra were measured on an ion-implanted region outside the disk, shown in Figure 3c (middle). The top spectrum in Figure 3c shows emission collected from the ion-implanted region in the disk. In this case, ZPL of $NV_{hh}$ overlaps with one of the cavity modes, demonstrating spectral coupling between the emitter and μ-Disk resonance. Given the multitude of optical modes present in a μ-Disk, we next summarize their measured quality factor (Q) as a function of wavelength. An average of ~2000 Q-factor is observed within the 1150-1250 nm wavelength range and covering all NV ZPLs. Note the decrease in Q at longer wavelengths, which is attributed to reduced optical confinement near the tapered edge, where radiation and scattering losses become more pronounced. A similar tendency is reproduced in the FDTD simulations. A local reduction in Q is also observed around 1175-1200 nm, which we ascribe to geometrical variations in mode confinement or possible interactions with higher-order modes.

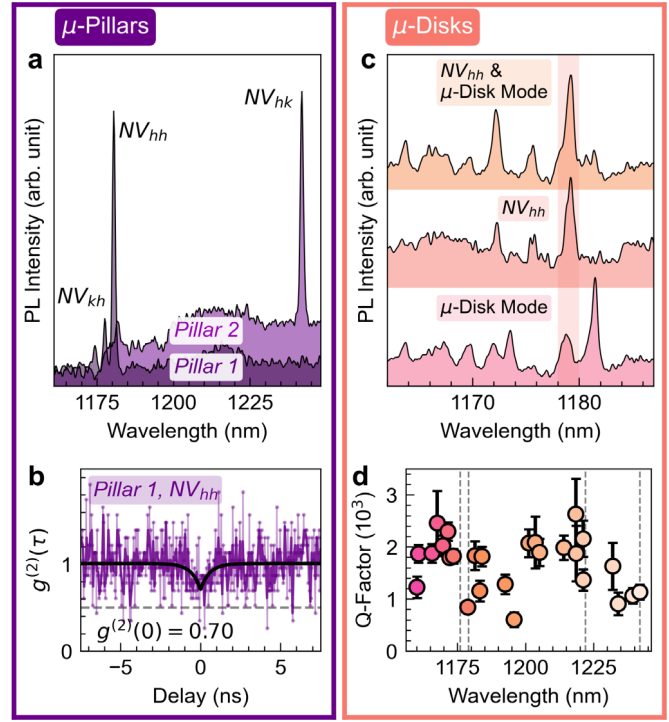

*Figure 3: Cryogenic temperature (12 K) optical characterization of NV centers in SiC μ-structures. (a) PL spectra of NV centers in two μ-Pillars. In Pillar 1, the ZPLs corresponding to the $NV_{kh}$ and $NV_{hh}$ are observed, while Pillar 2 exhibits ZPL associated with the $NV_{hk}$ orientation. (b) Second-order autocorrelation measurement from $NV_{hh}$ in Pillar 1. (c) PL spectra from a μ-Disk hosting NV centers. (d) Distribution of the μ-Disk Q-factors as a function of wavelength. Error bars are obtained from covariances of single Lorentzian fits.*

To further assess the performance of NV centres in the fabricated μ-structures and to quantify their performance for sensing applications we conduct optical and spin readout measurements under ambient conditions (295 K). Scanning confocal PL maps of the μ-Pillar arrays are shown in Figure 4a, where pillars appear as bright, localised emission sites with strong signal relative to the surrounding bulk. Pillar 1, highlighted in the map, was previously analysed under cryogenic conditions and is further characterised in this section.

The optical saturation in Figure 4b uses photon counts recorded by a superconducting single-photon detector (SSPD) and compares Pillar 1 to NV centers in bulk material (location indicated in Figure 4a). At any given optical excitation power, the μ-Pillar exhibits significantly higher count rates. For example, with an excitation power of 6 mW, Pillar 1 emits an average of $600(10^3)$ counts compared to $140(10^3)$ counts from bulk. Using this data we applied a saturation model (see Supplementary Section 4) that shows nearly identical saturation powers of 32.76 mW and 32.73 mW for Pillar 1 and bulk, yet, the extracted saturation intensities differ substantially, reaching $3845(10^3)$ counts for Pillar 1 and $916(10^3)$ counts for bulk. This ~4× enhancement highlights the benefit of improved photon collection from the directed emission cone of the pillar geometry.

The advantage of increased photon counts is also demonstrated in optical spin readout. Figure 4c compares ODMR spectra acquired from bulk and Pillar 1 for the same acquisition durations. While both datasets exhibit the expected $NV_{hh}$ and $NV_{kh}$ resonances at 1.33 GHz and 1.35 GHz, respectively, the bulk spectrum displays significantly higher noise compared to the clear features observed from Pillar 1. To quantify this improvement, each spectrum was fitted with a triple-Lorentzian model (defined in Supp. Section 4). For each resonance, the root-mean-square (RMS) uncertainty derived from the fit covariance matrix was extracted within the full width at half maximum (FWHM). For the $NV_{hh}$ and $NV_{kh}$ resonances, the extracted RMS values decrease from $0.66 \pm 0.09$ and $0.58 \pm 0.07$ (bulk) to $0.28 \pm 0.04$ and $0.24 \pm 0.03$ (Pillar 1), resulting in an approximate 2.4× reduction in spectral noise. This significant improvement has great implications for sensing capabilities of the NV center and will be discussed later in the manuscript. Additionally, the strong resonance at 1.296 GHz, attributed to a VV centre, similarly exhibits improved signal quality in the μ-Pillar structure. A full comparison of extracted fitting parameters is provided in Supplementary Section 4.

Moving on to the μ-Disks geometry, we first study the spatial PL shown in Figure 4d. A rectangular region at the top of the map corresponds to the area of focused ion irradiation, which partially overlaps with the fabricated μ-Disk to implant nitrogen in a localised area of the structure. Emission intensities from the irradiated disk region and a nearby bulk region are compared in Figure 4e. Due to the high overall defect density in this sample, particularly within the bulk, the recorded SSPD counts do not exhibit a clear saturation behaviour. Nevertheless, at all excitation powers, the μ-Disk region consistently produces higher photon count rates compared to bulk, despite its smaller volume. This indicates enhanced photon extraction arising from the disk geometry.

Increased photon collection is further reflected in spin readout performance. ODMR spectra acquired from the μ-Disk and bulk under identical excitation powers and acquisition durations demonstrate improved spectra in the fabricated structure. Using the same triple-Lorentzian fitting procedure and RMS uncertainty analysis described for the μ-Pillar, an approximate 1.4× reduction in spectral noise is observed for the NV resonances in the μ-Disk. Extracted parameters are summarised in Section 4 of the Supplementary Material. A weaker resonance at 1.395 GHz, attributed to a VV centre, is also present in the spectra and exhibits similarly improved signal quality within the μ-Disk structure.

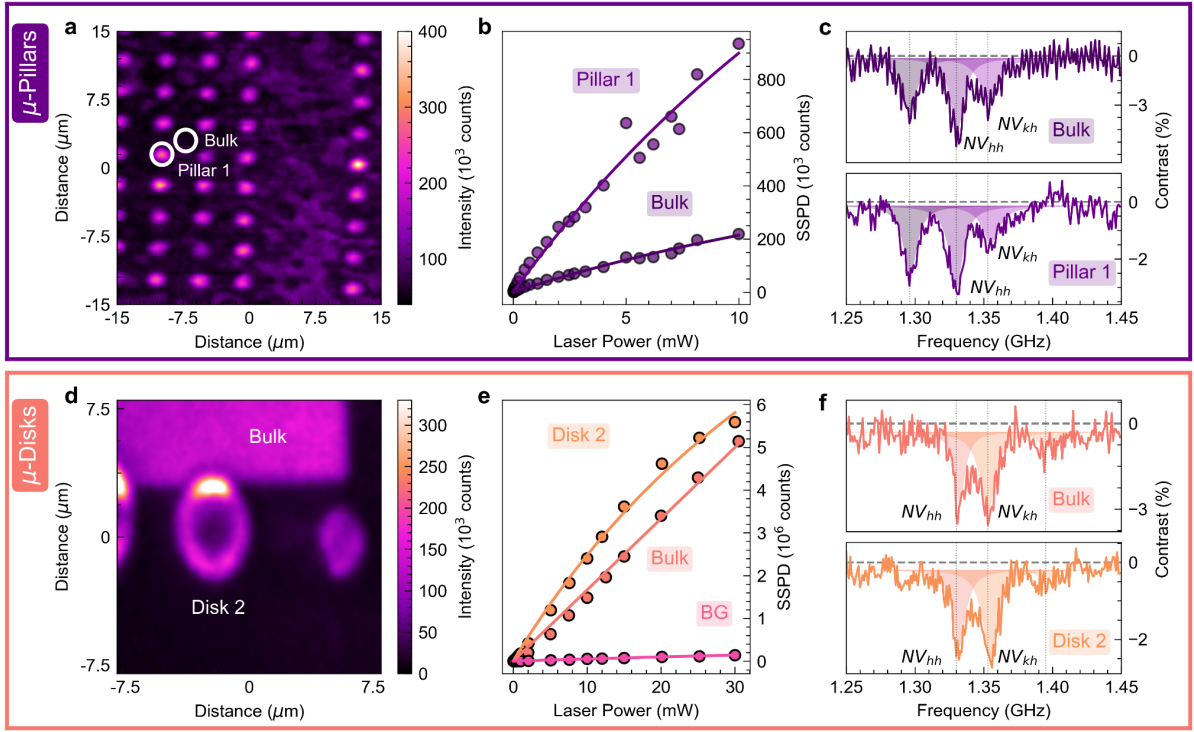

*Figure 4: Room-temperature optical characterization of NV centers in SiC µ-structures. (a) Confocal PL map of µ-Pillars using excitation power of 3 mW. (b) Comparison of photon counts from Pillar 1 and the bulk region. (c) ODMR spectra from bulk (top) and Pillar 1 (bottom). (d) Confocal image of µ-Disks using excitation power of 1 mW. (e) Comparison of photon counts from Disk 2, the bulk region with ion implantation, and the bulk region without ion implantation (BG). (f) ODMR spectra of the bulk region (top) and Disk 2 (bottom).*

Improvement in ODMR spectral quality has direct implications for the sensing capabilities of NV centers. When employing color centers as DC magnetic field sensors, their performance is commonly characterised using the sensitivity model[25,26]:

$$\eta_{DC} = \mathcal{P}_L \frac{h}{g_e \mu_B} \frac{\Delta\nu}{C\sqrt{R}} \quad (1)$$

Where $\mathcal{P}_L \approx 0.77$ is a constant associated with the Lorentzian line shape, $h$ is Planck's constant, $g_e = 2$ is the electron g-factor, $\mu_B$ is the Bohr magneton, $\Delta\nu$ is the ODMR linewidth, $C$ is the contrast, and $R$ is the detected photon count rate.

Despite the slight reduction in contrast for the $NV_{hh}$ transition in the µ-Pillar structure, likely due to variations in densities of signal and background emissions, enhanced PL signal still leads to improved DC magnetic field sensitivity, which scales inversely with the square root of the photon count rate. Therefore, Given the ~4× enhancement in collected photon counts at saturation, an approximate ≥2× improvement in sensitivity is expected. Considering the ODMR acquired in Figure 4c, at roughly half the saturation point, for the $NV_{hh}$ transition we record a sensitivity improvement from $13.01 \pm 1.76$ µT/$\sqrt{Hz}$ in bulk to $9.92 \pm 0.99$ µT/$\sqrt{Hz}$ in the µ-Pillar. These results once again demonstrate a scalable pathway towards improved sensing performance in SiC NV centres and other spin-active defect systems, such as the VV centre also observed in Figure 4.

**Conclusion**

In conclusion, we demonstrated coupling of 4H-SiC NV centers to photonic micro-structures, specifically µ-Pillars and µ-Disks. The µ-Pillar devices exhibited an approximately 4-fold enhancement in photon collection from the increased NA of the system. This allowed for the first observation of non-classical emission at cryogenic temperatures, while simultaneously reducing noise in ODMR spectra by a factor of 2.4. Consequently, the calculated sensitivity was improved by ~24 %. The µ-Disk structure exhibited resonance peaks ranging from 1150 nm to 1250 nm with an average Q factor of ~2000. Moreover, spectral overlap between the $NV_{hh}$ emission and the WGMs was observed. With further optimisation of the structure, WGMs of the µ-Disk can be designed to spectrally align with all four NV ZPLs. Lastly, we note that improvement in device fabrication and suppression of background emissions from VVs will further enhance collection efficiency and signal-to-noise ratio in the ODMR spectra. Overall, these results demonstrate the SiC NV center integration into µ-Pillars and µ-Disks geometries as a scalable platform for quantum technologies.

**Acknowledgments**

The authors acknowledge financial support from the Australian Research Council (CE200100010, FT220100053, and DP250100973, the Air Force Office of Scientific Research (FA2386-25-1-4044). We acknowledge access to NCRIS funded facilities and expertise at the ion-implantation Laboratory (iiLab), a node of the Heavy Ion Accelerator (HIA) Capability at the Australian National University.

**Competing Interests**

The authors declare no competing financial interest.

**Methodology**

*Sample Fabrication*

Production-grade silicon carbide wafers were commercially sourced from Wolfspeed and subsequently subjected to irradiation and nanofabrication processes. For the µ-Pillar geometry, samples were first implanted with $N^+$ ions using broad-beam irradiation at an accelerating voltage of 200 keV. Following implantation, the structures were defined using electron beam lithography and a hard metal mask. Structures were etched in a fluorine plasma environment, after which the mask was removed using a wet chemical etchant. µ-Disk structures were fabricated prior to implantation and subsequently irradiated using a focused ion beam operating at 30 keV.

*Optical Characterization*

All optical measurements were carried out using home-built scanning confocal microscope systems operating in reflection measurement. Samples were addressed using a 1050 nm continuous wavelength (CW) laser coupled into a single-mode optical fibre (Thorlabs) and out-coupled through a fibre launcher equipped with a double-achromatic lens. This ensured Gaussian beam profile for optimized excitation. Downstream optical components included a 1064 longpass dichroic mirror (Semrock) and an apochromatic 0.81 NA objective (Attocube LT-APO). The collected PL signal was coupled into a singlemode fiber (SMF-28) connected directly to a superconducting single photon detector (Scontel) or a spectrometer charge coupled detector (Princeton PyLoN IR) using 300 l/mm or 1200 l/mm gratings. Correlation measurements used a singlemode 50:50 beamsplitter fiber with coincidence counts analysed by a counter module (Swabian Instruments Timetagger 20).

Room temperature measurements were carried out in ambient conditions. For cryogenic measurements samples were cooled down to 25 K using a closed-loop helium cryostat (Attocube attoDRY 800) and kept at pressure of ~10-5 mbarr.

For ODMR measurements microwave excitation was provided by a radio-frequency signal generator (AnaPico APSIN 4010), with the output delivered to the sample via a suspended copper wire placed in close proximity to the structures. The microwave signal was amplified using a high-power amplifier (Mini-Circuits ZHL-16W-43-S+), while the frequency was swept within the 0.5~2.5 GHz resonance range, monitoring PL signal to obtain ODMR spectra.

**Data Availability Statement**

The data that support the findings of this study are available from the corresponding author upon reasonable request.